\documentstyle[aps,prl,twocolumn,psfig]{revtex}
\draft
\def\beq{\begin{equation}}
\def\eeq{\end{equation}}
\def\beqarr{\begin{eqnarray}}
\def\eeqarr{\end{eqnarray}}

\input epsf.tex

\begin{document}
\draft

\twocolumn[\hsize\textwidth\columnwidth\hsize\csname @twocolumnfalse\endcsname

\title{Core Strings and Flux Spreading Near Pinning Centers}
\author{D. J. Priour Jr$^{1,2}$, H. A. Fertig$^1$}
\address{$^1$Department of Physics and Astronomy, University of Kentucky, Lexington, KY40506-0055\\
	 $^2$Center for Computational Sciences, University of Kentucky, Lexington, KY40506-0045}

\date{\today}

\maketitle

\begin{abstract}

At the nucleus of a superconducting vortex is 
a small, circular core region where superconductivity
is destroyed.  Like an atomic nucleus, 
this core may become deformed, and such 
distortions can have important consequences
in non-equilibrium situations.
Using Ginzburg-Landau
theory, we have investigated this phenomenon for
vortices in the presence of artificial defects.
We show that when a vortex approaches 
the vicinity of a defect, an abrupt transition occurs in which the vortex
core develops a ``string'' extending to the defect boundary, while
simultaneously the supercurrents and associated magnetic
flux spread out and engulf the defect.
The energetics of stretching the string
determines the
pinning behavior of the vortex.
Experimental consequences of these strings are discussed.

\end{abstract}


]

%
%
One of the most important properties of superconductors
is their ability to carry currents without dissipation,
allowing them to generate large magnetic
fields.  Many superconductors allow fields to penetrate
in bundles of quantized magnetic flux, with associated whirlpools
of current known as vortices.  When these vortices are mobile, they
spoil the perfect conductivity that make superconductors so useful. 
The quest to increase the maximum dissipationless
current $J_{c}$ that a superconductor may carry has thus
fueled intense study of vortex pinning.

In recent years,   
pinning environments of artificially 
fabricated nanoscale defect arrays have been  
developed in hopes of better understanding and improving
the pinning properties of superconductors.  
Some of the earlier contributions have 
involved macroscopic measurements (e.g. $J_{c}$, magnetization) 
performed on periodic ``antidot'' arrays\cite{Harada,Baert2,MoshB96}. 
The antidot regions contain material which is
rendered non-superconducting. 
Pinning behavior may be studied in these periodic systems
using various imaging techniques\cite{Harada,Motton,Trapton,Field,Tona}. 
  

Much of the theoretical work on vortex pinning has 
employed numerical studies to focus\cite{Nori} 
on the behavior of large collections
of vortices under the influence of a driving force
(supercurrent). 
Such studies usually
employ simplified pinning potentials, in part to 
make possible simulations of large numbers of
vortices, but also because information about
pinning potentials at the microscopic level
is simply unavailable.  A few studies\cite{buzdin,mkrtch} 
have focused on energy scales
for pinning varying numbers of flux quanta to
the defects, as well as defect-vortex potentials
as derived from the London equation\cite{mos,cheng}.
However, the latter approach does not allow for
variation of the Cooper pair density, and in particular
cannot correctly treat the vortex core.
In the London approach,  vortex cores are usually assumed to
be rigid in shape, and interactions of vortices with their environments are
determined by the core position as well as the distribution 
of currents\cite{mos,cheng}.
Our results demonstrate that the vortex core in fact deforms dramatically
near an artificial defect: when the vortex center is sufficiently close
to the defect, a {\it string} of suppressed order parameter
develops from the vortex position to the pinning center edge, while
the currents and magnetic flux spread out over a large area
(see Fig. 2 below).  The pinning potential of the vortex turns out to be dominated
by the string when it is present, as we discuss below.

{\it Methods--}
   Our calculations focus on two dimensional arrays
of artificial pinning centers in the form of
holes in a bulk superconductor.
Our goal is to find the lowest energy state
of the system for
specified locations of a superconducting vortex;
from this we can construct a pinning
potential.  The appropriate
description of the superconducting state is in terms
of Ginzburg-Landau theory, which focuses on
a complex superconducting order parameter $\psi(\vec{r})$,
for which $\left| \psi \right|^{2}$ is proportional to the 
local density of superconducting electrons.   Unlike the 
London theory, Ginzburg-Landau theory is valid at scales
as small as the coherence length, $\xi$.
Written in terms of dimensionless variables, the
Ginzburg-Landau energy functional is

\begin{equation}
E_{GL} = \int \left[ \begin{array}{c} \left| \psi^{*} \left( 
\vec{\nabla}/i - \vec{A} \right) \psi \right|^{2} - 
\left| \psi \right|^{2} \\ + \frac{\kappa^{2}}{2} \left| \psi
\right|^{4} + B^{2}  
\end{array} \right] d^{3}x.
\label{Eq:eq1}
\end{equation}  
In Eq.~\ref{Eq:eq1}, $\vec{A}$ is the 
vector potential, and the magnetic field 
$\vec{B}(\vec{r}) = \vec{\nabla} \times \vec{A}$.
$\kappa \equiv \frac{\lambda}{\xi}$ is the Ginzburg parameter, the 
ratio of the magnetic penetration depth $\lambda$ 
and the coherence length.  In this work, we report on results 
obtained for $\kappa = 8$.  This choice of $\kappa$ is deep 
enough into the high $\kappa$ limit that, apart from scale factors 
in $E_{GL}$ and $\psi$, the results vary little
as $\kappa$ is increased.  

To analyze the behavior of the vortex near the pinning center,
we employ a mean field approach in which
one
minimizes $E_{GL}$ for a fixed vortex location to find $\psi$, $\vec{A}$,
and the current $\vec{J}$.  
Our strategy 
for calculating $\psi$ and $\vec{A}$ self-consistently involves first holding 
$\psi$ fixed at some initial guess,
and minimizing $E_{GL}$ with respect
to  $\vec{A}$ and $\vec{B}$.  Next, we fix $\vec{A}$ 
and $\vec{B}$ and minimize with respect to $\psi$.  
These steps are iterated until 
changes in the variables become negligible.  
We implement this self-consistent approach numerically by
dividing the unit cell into a fine lattice of small unit cells.  In this discrete
scheme, $\psi(\vec{r})$ is replaced by $\psi_{ij}$ with $ij$ specifying
a grid point on a square lattice, while $\vec{A}_{ij}$
and $\vec{J}_{ij}$ are defined on nearest-neighbor links
between the grid points.
Derivatives in Eq.~\ref{Eq:eq1} are replaced by the 
corresponding finite differences.  
The resulting theory is very similar to 
lattice gauge theories studied in particle physics.
To model the defect array as accurately as 
possible, one desires a fine grid; we find
that with a $128 \times 128$ mesh our results are well-converged
with respect to the discretization.  

To see how one minimizes $E_{GL}$ under the constraint of a
specified vortex location, it is 
useful to write the current $\vec{J}$ in  terms of the order parameter $\psi$ 
and the vector potential $\vec{A}$.  By minimizing $E_{GL}$  with respect
to the vector potential and employing a Maxwell equation one has 
\begin{eqnarray}
\label{Eq:eq2}
\vec{J} &=& \frac{1}{2} \left[ \psi^{*} \left( \vec{\nabla}/i - 
\vec{A} \right) \psi + \psi \left( -\vec{\nabla}/i - \vec{A} \right) \psi^{*} 
\right] \\
&=& \left| \psi \right|^{2} \left( \vec{\nabla} \phi - \vec{A} \right) . \nonumber
\end{eqnarray} 
In Eq.~\ref{Eq:eq2}, we have used for the order parameter $\psi = \left| \psi 
\right | e^{i\phi}$.  The familiar fluxoid quantization condition\cite{tinkham} arises from 
the requirement that the order parameter be single valued, i.e. $\oint \vec{\nabla} \phi \cdot
d \vec{s} = 2 \pi n_{v}$.  Hence, in terms of $\vec{J}$ and $\vec{A}$,  
\begin{eqnarray}
\label{Eq:eq3}
2 \pi n_{v}(ij) &=& \oint \left( \vec{J}/\left| \psi \right|^{2} \right) \cdot d\vec{s} + 
\oint \vec{A} \cdot d\vec{s}\\  
&=& \oint \left(  \vec{J}/\left| \psi \right|^{2} \right)\cdot d\vec{s} \nonumber
+ \Phi_{B} 
\end{eqnarray}
In the second part of Eq.~\ref{Eq:eq3}, $\Phi_{B}$ is the total magnetic 
flux passing through the area of the contour, which we conveniently 
choose to be the small unit cell associated with the grid point $ij$, 
while $n_{v}(ij)$ is the total number of ``fluxoid quanta'' 
contained in the contour of integration\cite{tinkham}. 
It is through $n_{v}$ that the vortex location(s)
in the full unit cell of the system may be fixed:
$n_{v}=0$ except at the grid points where we
wish to place a vortex, for which $n_{v}=1$. 
Armed  with knowledge of $|\psi|$ and some specified realization
of $n_{v}(ij)$, one can
solve for $\vec{J}$ and $\vec{A}$ via Eq.~\ref{Eq:eq3}.  Using the expression 
for the current given in Eq.~\ref{Eq:eq2}, 
one obtains $\vec{\nabla} \phi$; inserting $\vec{\nabla} \phi$ and $\vec{A}$ 
into Eq.~\ref{Eq:eq1} yields an expression depending only on $\left| \psi \right|$
and $\kappa$ which we minimize with respect to $\left| \psi \right|$ to obtain
 the order parameter modulus.  

We note that the above method can be generalized to the case of a
thin film superconductor.  We have performed some calculations for
such systems, and find that for antidot systems the results obtained
are quite similar to the large $\kappa$ results we report here.
This may be understood in terms of the effective magnetic penetration
depth for a thin film, $\lambda_{eff}
= \frac{\lambda^{2}}{d}$, which is typically much larger 
than the bulk value $\lambda$\cite{pearl}. 
This means that the energy stored in the magnetic field generated
by the supercurrents is quite small, so that the fact that the field 
varies as one moves out of the plane has little impact 
on the state of the system.  The resulting energy functional is thus 
nearly identical to the bulk three dimensional case, with columnar antidots and 
vortices.

\begin{figure}
\begin{center}
\centerline{\psfig{figure=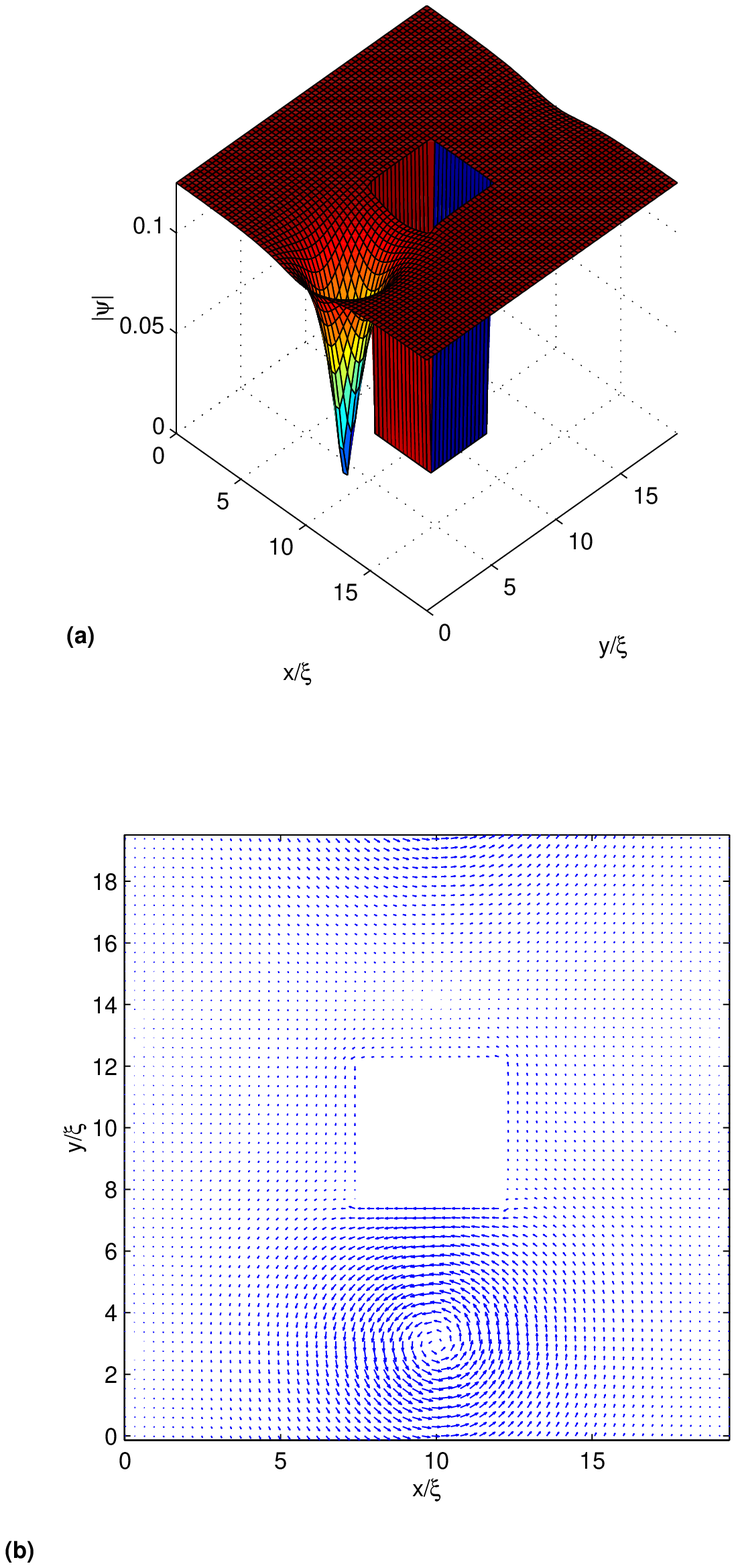,width=3.4in}}
\vspace{5mm}
\caption{A perspective plot of the order parameter modulus, $|\psi|$ (a) and current 
image (b) just prior to the transition.  Currents are localized about the vortex
core, which has a compact structure.} 
\label{Fig:fig2}
\end{center}
\end{figure}

\begin{figure} 
\begin{center}
\centerline{\psfig{figure=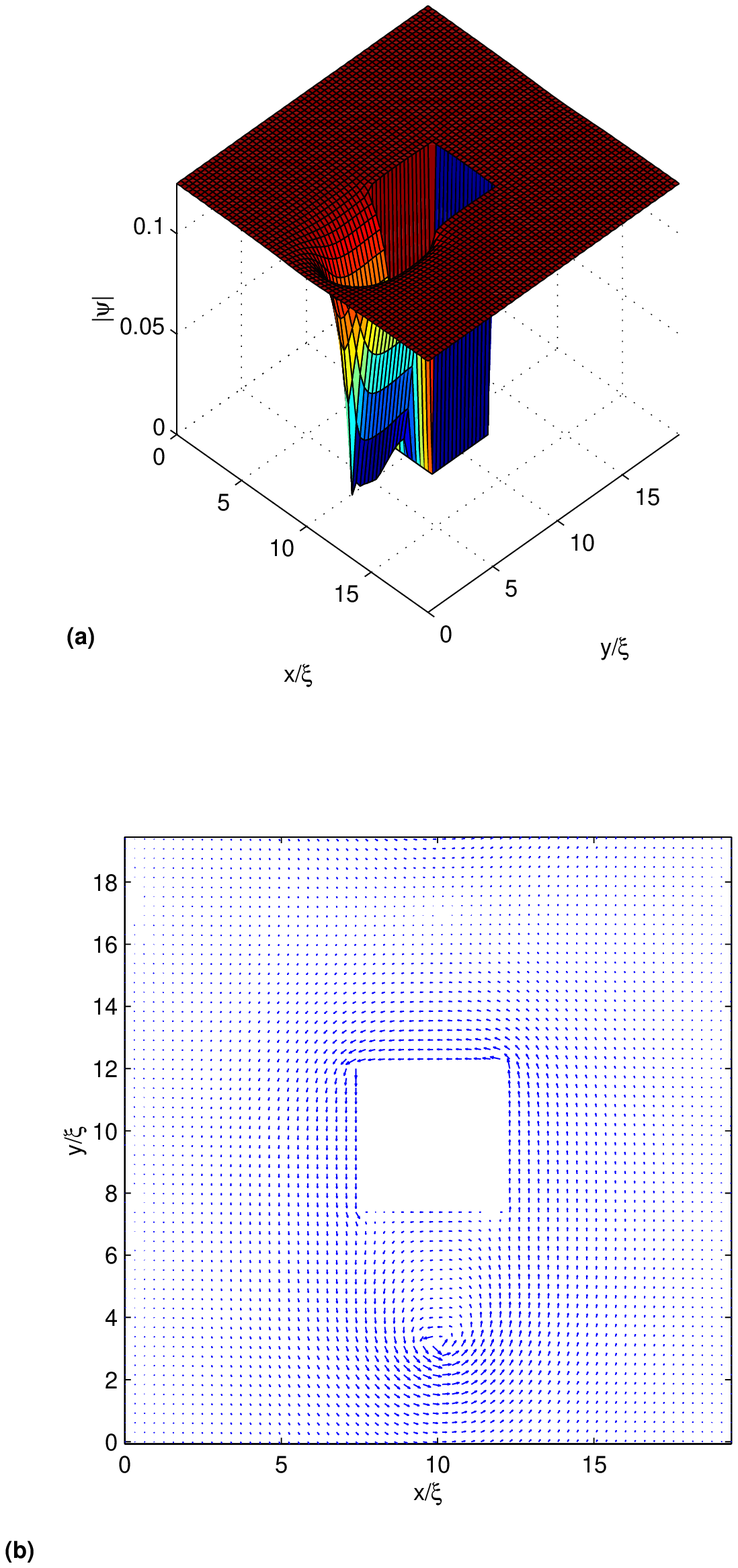,width=3.4in}}
\vspace{5mm}
\caption{Order parameter perspective plot (a) and currents (b) just after the transition.  
The currents circulate about the vortex-defect pair, and the vortex core has 
a string extending all the way to the defect edge.}
\label{Fig:fig3}
\end{center}
\end{figure}

{\it Results--}
At large vortex-defect separations, the core has 
the usual compact structure with supercurrents 
localized about it.  Fig.~\ref{Fig:fig2}
presents a perspective plot of $\left| \psi \right|$, as well as a vector plot of 
the currents.  The distances shown  
are in units of the coherence length, $\xi$.  
Our choice of a unit cell with side spanning 20 coherence lengths  
is typical of many of the nanoscale arrays studied experimentally.
As the flux quantum nears the defect 
edge, it eventually reaches a critical distance $d_{c}$ 
(typically several $\xi$, with precise value 
depending on the dot size and shape), where  
there is a sudden dramatic change.  Fig.~\ref{Fig:fig3} illustrates
the situation after the transition:
the vortex core has developed
a {\it string} extending from 
the flux quantum position to the defect edge; simultaneously the current now 
encircles the vortex-defect pair, and the magnetic flux created by these
currents spreads over a larger area. 
The string 
is energetically favorable because it allows the formerly 
dense current of the vortex to spread out (engulfing the 
defect in the process), thereby reducing the kinetic 
energy of the state. 
 

\begin{figure}
\begin{center}
\centerline{\psfig{figure=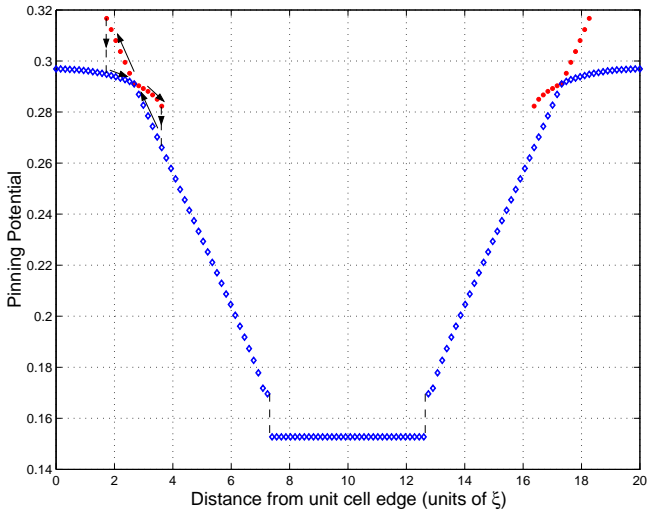,width=3.4in}}
\vspace{5mm}
\caption{Plot of the pinning potential showing broad linear 
regions and a gap in the potential at the defect boundary.  The 
horizontal axis measures the distance of the vortex from the 
edge of the unit cell.  Arrows and red circles indicate 
hysteresis in the calculation, a hallmark of a first order
transition.} 
\label{Fig:fig4}
\end{center}
\end{figure}

When a net supercurrent flows across a superconductor,
a Lorentz force is exerted on vortices; 
if these move in response energy
is dissipated.  In an artificial defect array, the 
driving force can be balanced by
a pinning force given by the gradient of $E_{GL}$
with respect to the vortex position.  Hence, it is appropriate to
regard $E_{GL}$ as the pinning potential. 
Fig.~\ref{Fig:fig4} shows this 
as a function of distance from the unit cell edge,
and it can be seen to have
three distinct regions.  For large separations, where 
the vortex core is compact, 
the pinning potential decreases relatively slowly.  
As the separation decreases it eventually crosses 
$d_{c}$ and
the core-string structure appears.  Hysteresis in the calculations, 
indicated with arrows, suggests that the transition is first 
order.  Below the transition one observes 
nearly linear  behavior of the pinning potential,  
suggesting that the
string carries an energy proportional to its length. 
The third region is  announced by a discontinuous
jump as the vortex is absorbed by 
the defect, followed by a perfectly flat region inside
the defect.  This  abrupt jump is due 
to a sudden transformation of the order parameter.  Immediately 
prior to absorption, constant Cooper pair density contours 
near the flux quantum have a semicircular profile.
As the vortex enters the defect, this ``semi-vortex'' vanishes,
removing a finite amount of energy
for an infinitesimal change in the vortex position.

{\it Experimental implications--} 
   The profile of the pinning potential shown in Fig.~\ref{Fig:fig4} suggests
a set of measurements one might perform to seek an experimental signature of 
the unique pinning phenomena discussed above.   
 One possible test would involve an AC driving
current of magnitude small enough to leave the vortices
pinned by the strings.  Such a current would allow the
vortices to ``rattle around'' in the linear part of the pinning well
illustrated in Fig.~\ref{Fig:fig4} and produce losses,
whereas a DC current of equal magnitude would be
dissipationless.  Upon lowering of the temperature
the vortices would be captured by defects, leading
to lossless supercurrents for both AC and DC driving
forces.  An observation of these effects would yield
indirect confirmation of the form of 
the pinning potential we find.  Something 
like this may recently have been observed\cite{lance2}. 
Another interesting possibility is that a unique thermal depinning 
may occur as the temperature is increased in 
the regime of linear pinning.  The presence of 
a string 
suggests that this may carry entropy at finite
temperature, much as is the case of polymers.
This entropy is proportional to the string length
and temperature, and at high enough temperatures
may overwhelm the energy
per unit length found in our mean-field
calculations.  In analogy with polymer behavior \cite{mrstring},
this leads to unbounded growth of the string and
effective depinning of the vortex.
However, it is not clear whether the string remains sufficiently
well-defined at the temperatures necessary for
proliferation that the polymer analogy remains valid up to the transition.
Further research into this possibility is currently underway.  
  
\begin{figure}
\begin{center}
\centerline{\psfig{figure=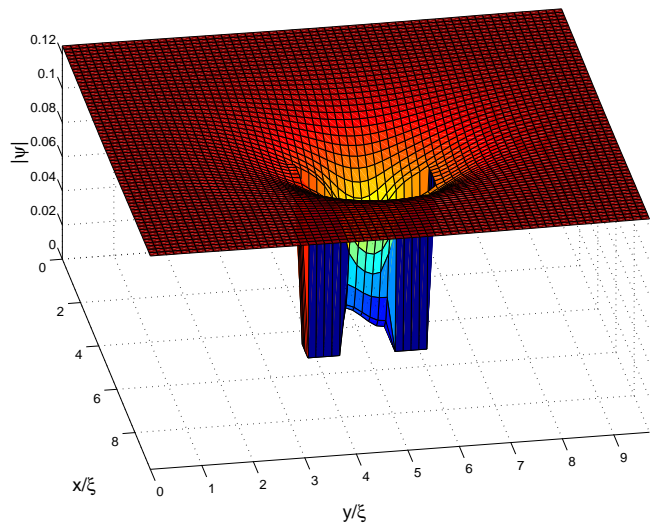,width=3.4in}}
\vspace{5mm}
\caption{Image of $|\psi|$ depicting a string connecting two defects 
in an experimentally realizable situation.}  
\label{Fig:fig5}
\end{center}
\end{figure}

Ultimately the best confirmation of our results would involve
imaging of the string.  
An interesting
possibility in this context is to fabricate arrays with
{\it two} defects per unit cell in close proximity.  When one magnetic
flux quantum passes through each unit cell,
we have found solutions 
in which a string develops between the defects.  
 Fig.~\ref{Fig:fig5} is an
image of $|\psi|$ for an equilibrium configuration in a typical case.  
The existence of a string between defects opens
the possibility of detecting these objects at low temperatures
and without driving currents,  
simplifying the relevant experimental conditions.

{\it Summary--}
     In the context of the Ginzburg-Landau theory, we have given a detailed 
treatment of the microscopic aspects of pinning phenomena in nanoscale periodic 
arrays.  Strikingly, we see an apparent first order 
transition involving the creation of a string connecting the vortices to
the defect,
and an accompanying abrupt transformation of the supercurrent
and the magnetic field it generates. 
The string configuration leads 
to a region of linear pinning.
Absorption of the vortex by the antidot is marked by a jump
in the pinning potential.  Various aspects of the pinning
potential should be observable
in experiment.

{\it Acknowledgments--}
The authors would like to thank L. E. DeLong, S. B. Field,
and J. B. Ketterson for useful discussions.
This work was supported by NSF Grant No. DMR-9870681
and DMR-0108451.

\end{document}